\documentclass[floatfix,aps,prb,twocolumn,showpacs,preprintnumbers,amsmath,amssymb]{revtex4}
\usepackage{epsfig}
\usepackage{graphicx}% Include figure files
\usepackage{dcolumn}% Align table columns on decimal point
\usepackage{bm}% bold math

\begin{document}
\title{The effect of morphology on the superconductor-insulator transition in 1-D nanowires}
\author{A. T. Bollinger, A. Rogachev, M. Remeika, and A. Bezryadin}
\affiliation{Department of Physics, University of Illinois at
Urbana-Champaign, Urbana, IL 61801-3080}
\date{\today}

%%%%%%%%%%%%%%%%%%%%%%%%%%%%%%%%%%%%%%%%%%%%%%%%%%%%%%%%%%%%%%%%%
%          ABSTRACT                                             %
%%%%%%%%%%%%%%%%%%%%%%%%%%%%%%%%%%%%%%%%%%%%%%%%%%%%%%%%%%%%%%%%%
\begin {abstract}
We study the effect of morphology on the low temperature behavior
of superconducting nanowires of length $\approx$100 nm. A
well-defined superconductor-insulator transition (SIT) is observed
only in homogenous wires, in which case the transition occurs when
the normal resistance is close to $h/4e^2$. Inhomogeneous wires,
on the other hand, exhibit a mixed behavior, such that signatures
of the superconducting and insulating regimes can be observed in
the same sample. The resistance versus temperature curves of
inhomogeneous wires show multiple steps, each corresponding to a
weak link constriction (WLC) present in the wire. Similarly, each
WLC generates a differential resistance peak when the bias current
reaches the critical current of the WLC. Due to the presence of
WLCs an inhomogeneous wire splits into a sequence of weakly
interacting segments where each segment can act as a
superconductor or as an insulator. Thus the entire wire then shows
a mixed behavior.
\end {abstract}


\maketitle

%%%%%%%%%%%%%%%%%%%%%%%%%%%%%%%%%%%%%%%%%%%%%%%%%%%%%%%%%%%%%%%%%
%          INTRODUCTION                                         %
%%%%%%%%%%%%%%%%%%%%%%%%%%%%%%%%%%%%%%%%%%%%%%%%%%%%%%%%%%%%%%%%%
Evidence for a superconductor-insulator quantum phase transition
(SIT) in one-dimensional (1D) wires has been found in a number of
experiments.\cite{Dynes, Bezryadin} Yet, other studies have
demonstrated a crossover, as opposed to an SIT, in thin wires
where superconductivity disappears gradually, as diameter is
reduced, presumably due to an increasing number of quantum phase
slips (QPS).\cite{Giordano, Lau, Lau2} Thus the existence and
possible origins of superconductor-insulator transitions in 1D
remain important open problems. In particular it is not known how
the SIT depends on the morphology of nanowires.

In two-dimensional system, for example, the crucial role of
morphology (i.e. granularity) on the SIT is well
known.\cite{Frydman, Garno, Finkelstein} For uniform films, as the
film thickness is reduced, a reduction of the critical temperature
is observed while the superconducting transition remains sharp.
The SIT occurs when the square resistance of the film reaches a
critical value close to the quantum resistance $R_Q$=$h/4e^2$=6.5\
\nolinebreak k$\Omega$.\cite{Goldman} In nonhomogeneous (granular)
films, on the other hand, a reduction of the film thickness
results in a {\em crossover} between superconducting and
insulating regimes, with very broad resistive transitions in the
thinnest superconducting samples.\cite{Frydman, Dynes2, Frydman2}
Our goal here is to determine the morphological requirements for
{\em 1D\ \nolinebreak nanowires} under which an SIT can occur.

In this Communication, we present a comparative study of
homogeneous and inhomogeneous nanotube-templated wires and find a
qualitatively different behavior at low temperatures. Homogeneous
samples (of length $\approx$100\ \nolinebreak nm) show an SIT
which occurs when the wire's normal resistance is close to $R_Q$,
confirming previous results where different nanotubes were
utilized as substrates.\cite{Bezryadin} Inhomogeneous wires, on
the other hand, exhibit a mixed behavior displaying properties of
superconductors and insulators at once. Such samples frequently
show multiple steps in the resistive transitions but no resistive
tails typical of QPS.\cite{Giordano, Lau, Lau2} We propose a model
which regards the inhomogeneous wire as a sequence of weak link
constrictions (WLC) connected in series. Each WLC has a certain
dissipative size and corresponding normal resistance, depending on
which the WLC can be either superconducting or insulating.
Homogeneous (and short enough) wires do not show such mixed
behavior because each wire acts as a single WLC.

\begin{figure}[b]
\begin{center}
\epsfig{file=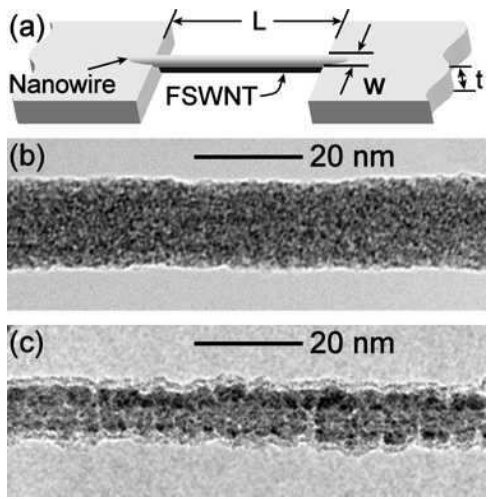} \caption{\label{fig:TEM} (a) Sample
configuration. The sample consists of two coplanar films
seamlessly connected by a thin wire. (b) A TEM micrograph of a
typical uncoated wire formed on a nanotube template. The MoGe
thickness is 8 nm. (c) A TEM micrograph of a Si-coated MoGe wire.
Sputtered MoGe thickness is 4.5 nm and Si thickness is 2 nm.}
\end{center}
\end{figure}

%%%%%%%%%%%%%%%%%%%%%%%%%%%%%%%%%%%%%%%%%%%%%%%%%%%%%%%%%%%%%%%%%
%          FABRICATION AND MEASUREMENTS                         %
%%%%%%%%%%%%%%%%%%%%%%%%%%%%%%%%%%%%%%%%%%%%%%%%%%%%%%%%%%%%%%%%%
The wires were fabricated using a molecular templating
technique,\cite{Bezryadin} in which a single-walled carbon
nanotube was suspended over a 100 \nolinebreak nm wide trench
etched into a multilayered Si/SiO$_2$/SiN substrate. Unlike
previous studies,\cite{Bezryadin, Lau} here we use {\em
fluorinated} single-wall nanotubes (FSWNT), which are known to be
insulating.\cite{Margrave} The substrate with suspended nanotubes
was then sputter-coated\cite{Sputtering} in one of two ways: (i)
with amorphous\cite{Graybeal} Mo$_{0.79}$Ge$_{0.21}$ or (ii) with
a slightly thinner amorphous Mo$_{0.79}$Ge$_{0.21}$ film followed
(in the same vacuum cycle) by a 2 \nolinebreak nm Si film. Contact
pads were defined using photolithography, followed by (i) wet
etching in H$_2$O$_2$ for MoGe or (ii) reactive ion etching
followed by wet H$_2$O$_2$ etching for Si-coated samples. While
both processes produce samples with the same geometry (Fig.\
\nolinebreak \ref{fig:TEM}a) and similar dimensions (Table\
\nolinebreak \ref{tab:table}), these two fabrication methods
result in different wire morphologies as revealed by TEM imaging
and confirmed by transport measurements (see below). The first
process results in homogeneous nanowires (Fig.\ \nolinebreak
\ref{fig:TEM}b), which we will refer to as ``bare'' wires. The
second method yields inhomogeneous wires, which will be called
``Si-coated'' wires (Fig.\ \nolinebreak \ref{fig:TEM}c). Although
these two fabrications methods give qualitatively different wires,
the exact mechanism of this is not clear. The most probable
explanation is that Si introduces some surface tension that leads
to the creation of grains. On the other hand, if exposed to air,
the outer layer of MoGe wires oxidizes unless coated by a
protective film, in our case Si. To ensure that the conducting
cores of bare wires have similar dimensions as the cores of the
Si-coated wires, the bare wires must be made thicker to compensate
for the expected surface oxidation. Since thicker wires are always
considerably more homogeneous than thin ones, the bare wires
exhibit a much higher degree of homogeneity.

\begin{table}
\caption{\label{tab:table}Parameters of nanowires. Samples A1-A12
are bare MoGe wires and B1-B7 are Si-coated wires. The normal
resistance of nanowires ($R_N$) was determined from $R(T)$ curves,
as indicated by an arrow for sample A1 in Fig.\ \ref{fig:RT}a.
Wire lengths ($L$) and widths ($w$) were measured with an SEM.
Here $t$ is the sputtered thickness of MoGe.}
\begin{ruledtabular}
\begin{tabular}{ccccc|ccccc}
\ & $R_N$ & $L$ & $w$ & $t$ & \ & $R_N$ & $L$ & $w$ & $t$\\
\ & (k$\Omega$) & (nm) & (nm) & (nm) & \ & (k$\Omega$) & (nm) & (nm) & (nm)\\
\hline
A1 & 2.39 & 99 & 21.4 & 8.5 & B1 & 4.85 & 126 & 15.4 & 6.0\\
A2 & 3.14 & 127 & 18.7 & 8.5 & B2 & 8.30 & 121 & 14.9 & 4.5\\
A3 & 3.59 & 93 & 16.8 & 8.5 & B3 & 9.52 & 139 & 12.1 & 4.5\\
A4 & 3.86 & 156 & 18.5 & 7.0 & B4 & 15.26 & 121 & 11.4 & 3.5\\
A5 & 4.29 & 188 & 20.8 & 7.0 & B5 & 16.31 & 123 & 13.8 & 4.0\\
A6 & 4.73 & 109 & 12.6 & 7.0 & B6 & 19.30 & 145 & 15.9 & 4.0\\
A7 & 5.61 & 116 & 11.6 & 7.0 & B7 & 44.83 & 115 & 13.0 & 2.0\\
A8 & 6.09 & 125 & 14.2 & 7.0\\
A9 & 8.22 & 105 & 10.6 & 5.5\\
A10 & 8.67 & 121 & 8.8 & 5.5\\
A11 & 9.67 & 140 & 10.8 & 5.5\\
A12 & 26.17 & 86 & 13.6 & 7.5\\
\end{tabular}
\end{ruledtabular}
\end{table}

\begin{figure}[t]
\begin{center}
\epsfig{file=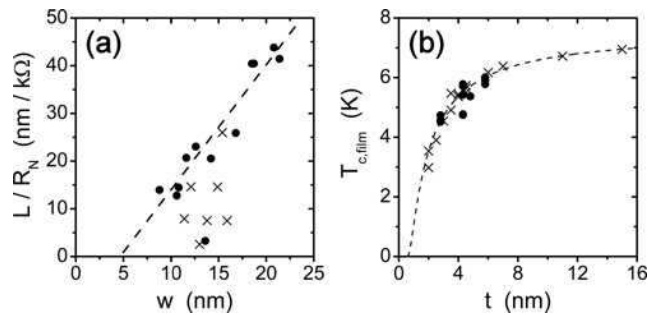} \caption{\label{fig:Oxidation} (a)
$L/R_N$ vs. $w$ for bare (circles) and Si-coated (crosses) wire
samples. The dashed line is a linear fit to the bare wire data,
$L/R_N = 2.6 \text{mS} (w - 4.6 \text{nm})$. (b) Film $T_c$ vs.
$t$ for bare (circles) and Si-coated (crosses) film samples. In
order for the data to agree, the bare MoGe film data have been
shifted from their actual position to the left by 2.7 nm, which
accounts for the thickness of the oxidized surface layer. The
dashed curve is a theoretical fit (Eq. (3) in Ref.\
\onlinecite{Oreg}).}
\end{center}
\end{figure}

The oxidized layer thickness (in bare wires) can be estimated as
follows (Fig.\ \nolinebreak \ref{fig:Oxidation}). For each sample,
the wire length, $L$, divided by the normal state resistance,
$R_N$, is plotted versus SEM measured width, $w$ in Fig.\
\nolinebreak \ref{fig:Oxidation}a. The bare wire data (circles)
can be well approximated by a linear fit (dashed line), providing
evidence for their homogeneous structure. The fitting parameters
give an oxidized layer thickness $\approx$2.3\ \nolinebreak nm
(half of the fit's x-axis intercept value) and the resistivity for
MoGe $\rho$$\approx$180\ \nolinebreak $\mu\Omega\text{-cm}$
(obtained from the fit's slope and the wire's average thickness,
$d_{ave}$$\approx$4.7\ \nolinebreak nm, estimated under the
assumption that the top 2.3\ \nolinebreak nm of the sputtered MoGe
is oxidized), in agreement with published values.\cite{Bezryadin,
Graybeal} No reasonable linear fit could be obtained for the
Si-coated wires (crosses), further indicating their inhomogeneous
structure. In Fig.\ \nolinebreak \ref{fig:Oxidation}b we compare
the $T_c$'s of bare MoGe films and Si-coated MoGe films (circles
and crosses, respectively) plotted versus their thicknesses. By
shifting the data for bare films by 2.7\ \nolinebreak nm to the
left the two families of data points overlap (Fig.\ \nolinebreak
\ref{fig:Oxidation}b). It is therefore concluded that the oxidized
layer thickness is 2.7\ \nolinebreak nm, similar to the above
estimate.

Voltage vs. current measurements, $V(I)$, were performed by
current biasing the sample through a large ($\sim$1 \nolinebreak
M$\Omega$) resistor. Zero-bias resistance, $R(T)$, was obtained
from the slope of the linear part of the $V(I)$ curves as
temperature was varied. Similar to Ref.\ \nolinebreak
\onlinecite{Rogachev}, transport measurements were performed in
$^4$He or $^3$He cryostats equipped with rf-filtered leads. The
differential resistance vs. bias current, $dV(I)/dI$, was measured
using an AC excitation on top of a DC current offset generated by
a low-distortion function generator (SRS-DS360), again in series
with a ($\sim$1 \nolinebreak M$\Omega$) resistor.

%%%%%%%%%%%%%%%%%%%%%%%%%%%%%%%%%%%%%%%%%%%%%%%%%%%%%%%%%%%%%%%%%
%          EXPERIMENTAL RESULTS AND DISCUSSION                  %
%%%%%%%%%%%%%%%%%%%%%%%%%%%%%%%%%%%%%%%%%%%%%%%%%%%%%%%%%%%%%%%%%
\begin{figure}[b]
\begin{center}
\epsfig{file=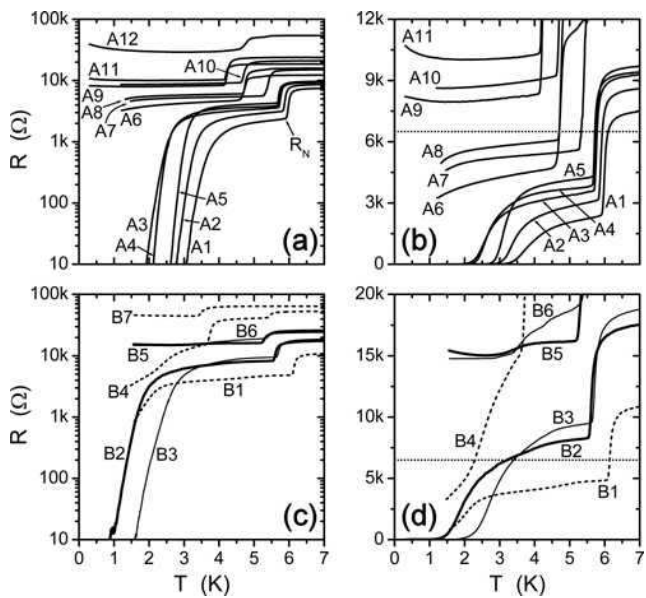} \caption{\label{fig:RT} Resistance vs.
temperature curves are shown for bare wires, which are homogeneous
(a and b), and Si-coated wires, which are inhomogeneous (c and d).
Figures (b) and (d) show the data on a linear scale for clarity.
The horizontal dotted lines indicate the $R_Q$ level at which the
SIT occurs in homogeneous samples.}
\end{center}
\end{figure}

Zero-bias resistance vs. temperature measurements are compared for
homogeneous (Figs.\ \nolinebreak \ref{fig:RT}a,b) and
inhomogeneous (Figs.\ \nolinebreak \ref{fig:RT}c,d) nanowires.
First note that the rightmost superconducting transition observed
on {\em all} samples is due to the electrodes (i.e. thin MoGe
films connected to the ends of the wire). Below the temperature of
this transition ($T_{c,film}$) only the wire contributes to the
measured resistance. Therefore, in what follows we will only be
interested in temperatures $T$$<$$T_{c,film}$. We first discuss
the family of $R(T)$ curves for bare wire samples, presented in
Fig.\ \nolinebreak \ref{fig:RT}a, which show a clear dichotomy. In
the log-linear representation, samples A1-A8 all have negative
curvature ($d^2\text{log}(R)/dT^2$$<$0) with more than half of
these samples reaching immeasurably low resistances. Therefore, we
refer to these samples as superconducting. Although samples A6,
A7, and A8 do not reach such low resistances, their negative
curvature as well as their decreasing resistance with decreasing
temperature (Fig.\ \nolinebreak \ref{fig:RT}b) lead us to believe
they too are superconducting. Due to their smaller widths we
assume that their critical temperatures are suppressed\cite{Oreg}
such that they do not go through their full superconducting
transition within the temperature range studied. Samples A9-A12
show only the transition due to the film electrodes. Their
curvature is always positive (for $T$$<$$T_{c,film}$) with
increasing resistances as temperature is decreased. Consequently
we consider these samples to be insulating. The insulating regime
can be either due to a suppression of $T_c$ to zero\cite{Oreg} or
due to a proliferation of QPS\cite{Zaikin} (or both). Observation
of these two regimes indicates that a superconductor-insulator
transition occurs in this family of nanowires, confirming previous
results.\cite{Bezryadin} As illustrated in Fig.\ \nolinebreak
\ref{fig:RT}b, the transition from superconducting to insulating
behavior occurs when the wire's $R_N$$\approx$$R_Q$. Therefore it
can be suggested that the $R_N$ of the wire is the parameter that
controls the SIT. However, since wire length was not varied over a
wide range we cannot exclude the possibility that the
cross-sectional area of the wire is the true control parameter.

Measurements presented in Fig.\ \nolinebreak \ref{fig:RT}
demonstrate that bare and Si-coated wires are qualitatively
different. In the Si-coated wires the $R(T)$ curves are less
predictable, show multiple steps or ``humps'' (Figs.\ \nolinebreak
\ref{fig:RT}c,d), and in some cases exhibit a mixed behavior that
does not always allow them to be clearly identified as either
superconducting or insulating. For example, sample B4 shows three
steps on the $R(T)$ curve and a tail at the lowest temperature,
which could be the beginning of the fourth transition. The first
step, as always, reflects the superconducting transition in the
electrodes, but the other steps correspond to some weak links in
the wire itself. Some Si-coated wires exhibit both positive and
negative curvature at different temperatures. Such mixed behavior
is seen in samples B5 and B6, with their resistance initially
dropping as in superconducting wires (i.e. with a negative
curvature) but then starting to increase as in insulating wires.
The conclusion is that no clear SIT at $R_N$$\approx$$R_Q$ can be
observed in the family of non-homogeneous wires.

\begin{figure}[t]
\begin{center}
\epsfig{file=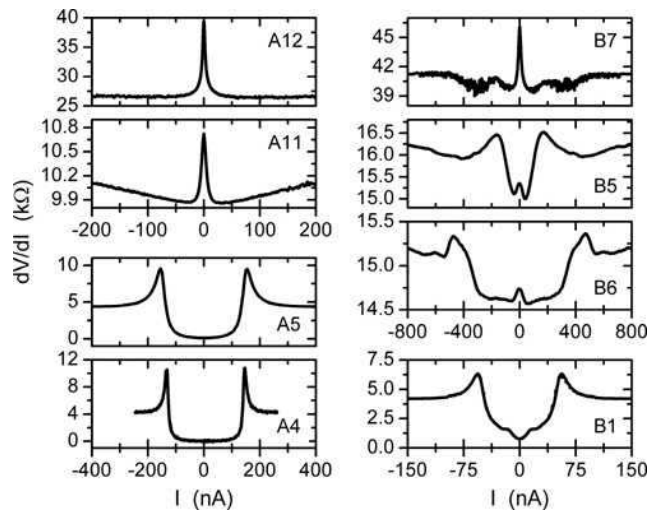} \caption{\label{fig:dVdI} Differential
resistance as a function of bias current for various bare (left
column) and Si-coated (right column) samples. Measurement
temperatures are 0.3 K for samples A11 and A12, 2.8 K for sample
A5, 2.2 K for sample A4, and 1.5 K for samples B1, B5, B6, and
B7.}
\end{center}
\end{figure}

A clear difference between bare and Si-coated wires is also found
in the $dV(I)/dI$ measurements (Fig.\ \nolinebreak
\ref{fig:dVdI}). Bare MoGe wires (left column) again show one of
only two distinct types of behavior. The superconducting wires
exhibit a large minimum centered around zero-bias (e.g. A4 and A5)
and peaks at the bias current which is equal to the critical
current of the nanowire. The insulating samples show a single
zero-bias resistance peak (e.g. A11 and A12). This is in agreement
with previous experiments.\cite{Bezryadin}

The Si-coated MoGe wires, on the other hand, frequently show
various combinations of such types of behavior. For example,
samples B5 has a clear resistance depression in the range $\pm$150
\nolinebreak nA with the critical current peaks at the limits of
this depression, and, in addition to this, a narrow zero-bias
resistance peak. Such mixed behavior indicates that some parts of
the wire are superconducting while some other parts act like
insulating wires. Sample B6 also shows similar results. This fact
again confirms that these wires are not homogeneous.

The properties of inhomogeneous wires can be understood by
assuming that they contain a sequence of independent weak link
constrictions, each surrounded by a dissipative region, the size
of which determines the normal resistance of the WLC. Since the
free energy barriers for phase slips are different in each such
WLC and the constrictions are connected in series it is clear that
$R(T)$ curves should show crossings\cite{Lau2} and multiple steps
(such as for sample B4 in Fig.\ \nolinebreak \ref{fig:RT}c).
Moreover some links can be insulating if their normal resistance
is large enough, in the same sense as homogeneous insulating
wires. For example sample B5 shows a mixed behavior that is
explained by assuming it contains two WLC - one superconducting
and one insulating. The $R(T)$ curve then should show one
superconducting hump and a tail with increasing resistance at
lower temperatures, typical of insulating wires. Likewise, a
single superconducting feature and an insulating zero-bias peak
should be observed in the $dV(I)/dI$ curve. These characteristics
are indeed observed (Fig.\ \nolinebreak \ref{fig:RT}d and Fig.\
\nolinebreak \ref{fig:dVdI}). In some wires all WLC can be
superconducting, as for example in sample B1 (and possibly B2, B3
and B4). In the case of B1, two humps in the $R(T)$ curve (at
$T$$\approx$5\ \nolinebreak K and $T$$\approx$2\ \nolinebreak K)
and two superconducting peaks on $dV(I)/dI$, corresponding to two
critical currents (at $I$$\approx$15\ \nolinebreak nA and
$I$$\approx$60\ \nolinebreak nA), are observed, indicating that
two superconducting WLCs are present. Finally, since each
insulating WLC produces a single $dV(I)/dI$ peak located at zero
current, the wires with a few insulating WLCs should nevertheless
exhibit only one peak at zero bias, as could be the case for
sample B7 in Fig.\ \nolinebreak \ref{fig:dVdI}.

All measured homogeneous samples act as superconductors if they
satisfy $R_N$$<$$R_Q$ and as insulators otherwise (Fig.\
\nolinebreak \ref{fig:RT}b). Inhomogeneous wires frequently
violate this condition, as for example sample B2 (Fig.\
\nolinebreak \ref{fig:RT}d). This can be explained by the WLC
model and by assuming that each WLC can be either insulating or
superconducting depending on its own normal resistance. The sample
B2 has shown two distinguishable peaks on its $dV(I)/dI$ curve
(not shown), corresponding to two superconducting WLCs with
slightly different critical currents. Since this sample's normal
resistance is 8.30\ \nolinebreak k$\Omega$ and since two similar
WLCs are detected in this wire, it can be estimated that the
normal resistance of each WLC is $\approx$4.15\ \nolinebreak
k$\Omega$. Thus each WLC is superconducting and hence the entire
wire is superconducting as well.

Since each WLC is manifest by a jumpwise increase of the sample
resistance at the WLC's critical current, it is reasonable to
associate WLCs with phase slip centers\cite{Skocpol,Tidecks}
(which are usually positioned at weak spots along the wire). The
normal resistance of a WLC is then the resistance of the
dissipative region or the region which is populated by
quasiparticles generated by the phase slip center. This dimension
is the quasiparticle diffusion length,
$\Lambda_Q$$\approx$$(D\tau_E)^{1/2}$$\approx$100\ \nolinebreak
nm, where $D$$\approx$$10^{-4}$\ \nolinebreak m\nolinebreak$^2$s
is the diffusion constant and $\tau_E$$\approx$$10^{-10}$\
\nolinebreak s is the inelastic scattering
time.\cite{Skocpol,Tidecks,Graybeal2} If constrictions are present
in the wire, $\Lambda_Q$ is further reduced and equals the
distance between the constrictions. Our short homogeneous wires
are not longer than $\Lambda_Q$, so each wire acts as a single
WLC. These wires therefore show a clear SIT. Inhomogeneous wires
have more than one WLC (i.e. will develop more than one phase slip
center under strong current bias) and consequently show a mixed
behavior.

In summary, we have found that the morphology of superconducting
nanowires has a strong effect on the observed
superconductor-insulator transition. A clear SIT is only found in
homogeneous wires whereas inhomogeneous samples show a smeared
transition and mixed behavior. In the later case, the results are
understood by the assumption that inhomogeneous wires are composed
of weakly coupled sections connected in series. Each section
exhibits either superconducting or insulating behavior while the
entire wire shows a mixed behavior.

This work is supported by NSF CAREER Grant No. DMR 01-34770 and by
the Alfred P. Sloan Foundation. Some fabrication was performed at
CMM-UIUC supported in part by DOE grant DEFG02-96-ER45439.

%%%%%%%%%%%%%%%%%%%%%%%%%%%%%%%
% BIBLIOGRAPHY AND REFERENCES %
%%%%%%%%%%%%%%%%%%%%%%%%%%%%%%%

%%%%%%%%%%%%%%%%
% END DOCUMENT %
%%%%%%%%%%%%%%%%

\begin{thebibliography}{99}
\bibitem{Sputtering}
The sputtering was DC for MoGe and RF for the Si overlayer. The
sputterer was equiped with an LN$_2$ cold trap.
\end{thebibliography}
\end{document}